%% file: ACMheader.tex
\documentclass[sigconf]{acmart}

\usepackage{booktabs} 

\setcopyright{none}

\acmDOI{??}

\acmISBN{??}

\acmConference[ACSAC]{ACSAC}{December 2020}{Austin, USA}
\acmYear{2020}
\copyrightyear{2020}

\acmArticle{?}
\acmPrice{15.00}

\begin{document}

\title[Metrics for Calculating Significant Software Flaw Types]{\input{Title.tex}}


\author{Peter Mell}
\orcid{0000-0003-2938-897X}
\affiliation{%
  \institution{National Institute of Standards and Technology}
}
\email{peter.mell@nist.gov}

\author{Assane Gueye}
\affiliation{%
  \institution{Prometheus Computing}
}
\email{a.gueye@prometheuscomputing.com}

\renewcommand{\shortauthors}{P. Mell et al.}

\input{Abstract.tex}

%
%

\ccsdesc[500]{Applied computing~Electronic commerce}
\ccsdesc[500]{Applied computing~Digital cash}
\ccsdesc[300]{Security and privacy}

\keywords{keyword1, keyword2, keyword3}

\maketitle

\input{MCbody}

\bibliographystyle{ACM-Reference-Format}
\bibliography{MCbody}

\end{document}

%% file: Title.tex
A Suite of Metrics for Calculating the Most Significant Security Relevant Software Flaw Types

%% file: Abstract.tex
\begin{abstract}

The Common Weakness Enumeration (CWE) is a prominent list of software weakness types. This list is used by vulnerability databases to describe the underlying security flaws within analyzed vulnerabilities. This linkage opens the possibility of using the analysis of software vulnerabilities to identify the most significant weaknesses that enable those vulnerabilities. We accomplish this through creating mashup views combining CWE weakness taxonomies with vulnerability analysis data. The resulting graphs have CWEs as nodes, edges derived from multiple CWE taxonomies, and nodes adorned with vulnerability analysis information (propagated from children to parents). Using these graphs, we develop a suite of metrics to identify the most significant weakness types (using the perspectives of frequency, impact, exploitability, and overall severity).
\end{abstract}

%% file: MCbody.tex
\section{Introduction}
\label{Intro}

The Common Weakness Enumeration (CWE) \cite{wu2015they} \cite{martin2007CWE} is a prominent list of software weakness types. 
It is maintained by the MITRE corporation, funded by the United States (U.S.) government, and developed with the participation of 55 organizations. 
The CWE list contains 808 weaknesses organized by multiple views. Views are `hierarchical representations' of CWEs (i.e., taxonomies) serving different communities with different perspectives on the data.

A specific view, 1003, was created to support the labelling of publicly disclosed software vulnerabilities (`potential weaknesses within sources that handle public, third-party vulnerability information'). It contains 123 software flaws. 
The National Vulnerability Database (NVD) \cite{NVD} and other vulnerability databases and security tools use view 1003 to describe the underlying security flaws within analyzed vulnerabilities. 98.8\,\% of the 12\,760 fully analyzed vulnerabilities published by NVD in 2019 were able to be mapped to view 1003, demonstrating its applicability and coverage.

This linkage of vulnerability analysis to the view 1003 CWEs opens the possibility of using the NVD analysis of software vulnerabilities to identify the most significant weaknesses that enable those vulnerabilities. In this work, we accomplish this through creating mashup views combining the following resources:
\begin{itemize}
    \item the multiple primary taxonomies of the CWE (views 1003, 1000, 699, and 1008),
    \item the Common Vulnerabilities and Exposures (CVE) \cite{CVE} enumeration of publicly disclosed software vulnerabilities,
    \item the NVD mapping of CVEs to view 1003 CWEs, and
    \item the NVD measurements of each CVE using the Common Vulnerability Scoring System (CVSS) \cite{CVSSv3Spec} \cite{mell2006common}. This calculates the exploitability, impact, and overall severity of each CVE outside of any particular deployment context.
\end{itemize}

The result of creating mashups of this data are graphs that have CWEs as nodes. The edges between the nodes are extracted from the parent-child relationships between the multiple CWE taxonomies. And the nodes are labelled with CVE and CVSS information (propagated backwards along the edges). We apply to these graphs a suite of simple metrics that we developed to identify the most `significant' weakness types. We evaluate significance from multiple perspectives using metrics focused on the following areas: frequency, impact, and exploitability. In doing this we evaluate the CWEs in two distinct groups to take into account the varying levels of abstraction of the CWEs. 

We create most significant weakness lists for each metric for the CVE vulnerabilities published in 2019 (not provided due to space limitations). We then analyze the differences between these six lists (3 metrics * 2 sets of CWE types) using two algorithms for comparing differences between ordered lists (Kendall's Tau and the Spearman's footrule variant \cite{Rankings}). We find that different weaknesses tend to emerge as the most significant depending upon the perspective, the metric used, and the CWE type. Note that we use simple low level metrics for our perspectives. This is because there is no ground truth for aggregating those metrics; equations in security that aggregate simple metrics are often practically useful but less scientifically defensible.

Finally, we note that the CWE already has an official metric to identify the `most dangerous' CWEs. It aggregates both frequency and severity with severity itself being an aggregate metric combining exploitability and impact. We discover weaknesses with this official metric that leads to the under counting of certain CWEs. 


We recommend that software developers and creators of software bug finding tools use our approach to prioritize finding and eliminating these most significant weaknesses to reduce the number and severity of security related flaws in software.



\section{Background}
As mashup research, our approach combines multiple resources. These are briefly described and referenced here.

\subsection{Common Weakness Enumeration}
Our research is primarily focused on the Common Weakness Enumeration (CWE) \cite{martin2008CWE}, a `community-developed list of common software security weaknesses'. `It serves as a common language, a measuring stick for software security tools, and as a baseline for weakness identification, mitigation, and prevention efforts' \cite{CWE}. The 808 software weaknesses within the enumeration are referred to as CWEs where each is named CWE-X with X being some integer. Each CWE is characterized as either a class, base, variant, or compound. Classes are the highest level of abstraction, followed by bases, and then by variants. Compounds are relatively rare and are combinations of multiple bases and/or variants. In our work we evaluate classes separately from bases, variants, and compounds, given that the classes have a much higher level of abstraction.

Besides the CWE weaknesses, there are also 295 categories and 38 views. Confusingly, these are also considered CWEs; for simplification we use the name CWE to refer only to the weakness CWEs. The categories are used to organize the CWEs within select views (this is not used in our research). The views are hierarchical organizations of a subset of CWEs according to some perspective (essentially a taxonomy). The three primary taxonomies are the `Research Concepts' (view 1000), `Development Concepts' (view 699), and `Architectural Concepts' (view 1008). This last view, 1008, was not useful to our work and is not used because it doesn't provide a hierarchy of CWEs but instead uses the categories to group CWEs. The view 1003 designed for vulnerability databases, mentioned previously, is called `CWE Weaknesses for Simplified Mapping of Published Vulnerabilities View' and is the core data structure upon which our work builds.

\subsection{Common Vulnerabilities and Exposures}
The set of software vulnerabilities used for this research come from the Common Vulnerabilities and Exposures (CVE) program, maintained by the MITRE corporation. `CVE is a list of entries—each containing an identification number, a description, and at least one public reference—for publicly known cybersecurity vulnerabilities' \cite{CVE} \cite{baker1999CVE}.

\subsection{Common Vulnerability Scoring System}
The Common Vulnerability Scoring System `provides a way to capture the principal characteristics of a vulnerability and produce a numerical score reflecting its severity' \cite{CVSS}. It provides equations for calculating a vulnerability's base score (inherent risk outside of any particular environment), temporal score (changing risk over time), and environmental score (risk within a particular environment). We use the base score, which is composed of two sub-scores that calculate the exploitability and impact of a vulnerability. It is maintained by the Forum of Incident Response and Security Teams (FIRST). The detailed specification for CVSS version 3.1 is available at \cite{CVSSv3Spec}. 

\subsection{National Vulnerability Database}
The National Vulnerability Database (NVD) is `the U.S. government repository of standards based vulnerability management data' \cite{NVD}. It is maintained by the U.S. National Institute of Standards and Technology. We use its scoring of CVEs with CVSS scores and its mapping of the CVEs to view 1003 CWEs.


\section{Foundational Data Structures}
This section describes how we generate the foundational data structure used by our metrics to calculate the most significant security relevant software flaw types. We generate a directed acyclic graph (DAG) of CWEs that we will use to propagate CWE analysis data between the CWEs.

\subsection{View 1003 Graph}

\begin{figure}
\centerline{\includegraphics[scale=.22]{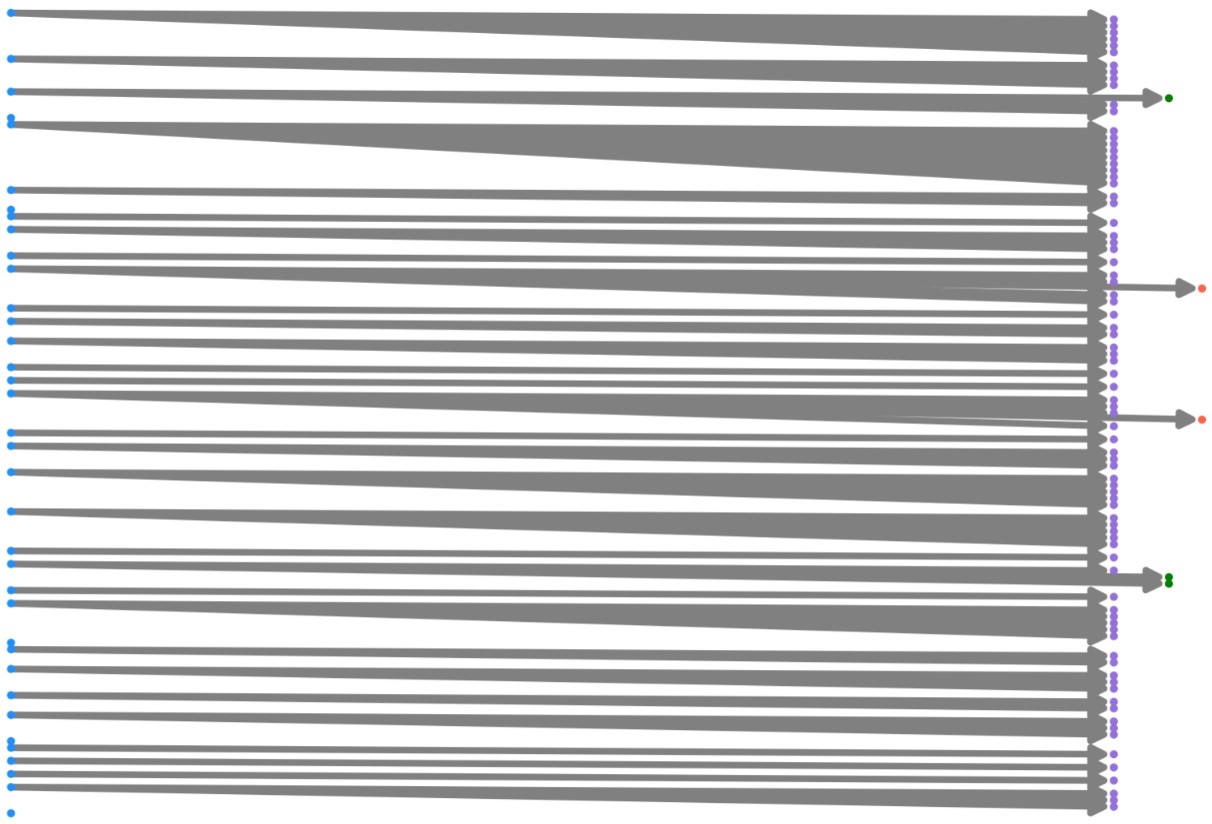}}
\caption{CWE View 1003 (123 nodes, 87 edges)}
\label{fig:Strict1003CWEGraph}
\end{figure}

We begin with the set of CWEs in CWE view 1003 since that is the set that was adopted by the NVD (and is the set identified by MITRE as most applicable to CVE vulnerabilities). We then form a graph of the view 1003 nodes through extracting the `ChildOf' relationships in the CWE view 1003 Extensible Markup Language (XML) file. Other kinds of relationships are provided in the XML file but we don't use them because none of them definitively indicate the parent child relationship needed to construct edges in our graph (for example, `CanPrecede'). The result is a rooted tree\footnote{A perfect tree structure is uncommon in weakness/vulnerability taxonomies. This encouraged us to explore possible missing relationships.} with the root being CWE 1003, the nodes at distance one from the root being classes, and the nodes at distance 2 being bases, variants, and compounds. We remove the root as we are only interested in the classes, bases, variants, and compounds. The resulting DAG has 123 nodes and 87 edges, shown in Figure \ref{fig:Strict1003CWEGraph}. On the left side are the 36 class nodes in blue. The majority of class nodes have edges to bases, variants, or compounds, but five do not. On the right side, the largest grouping of nodes in a single column in purple represents the 82 bases. Moved slightly to the right and in green are the 3 variants. Moved even farther to the right in orange are the 2 compounds. 

\subsection{Direct Edge Augmentation}
\label{directAug}

\begin{figure}
\centerline{\includegraphics[scale=.22]{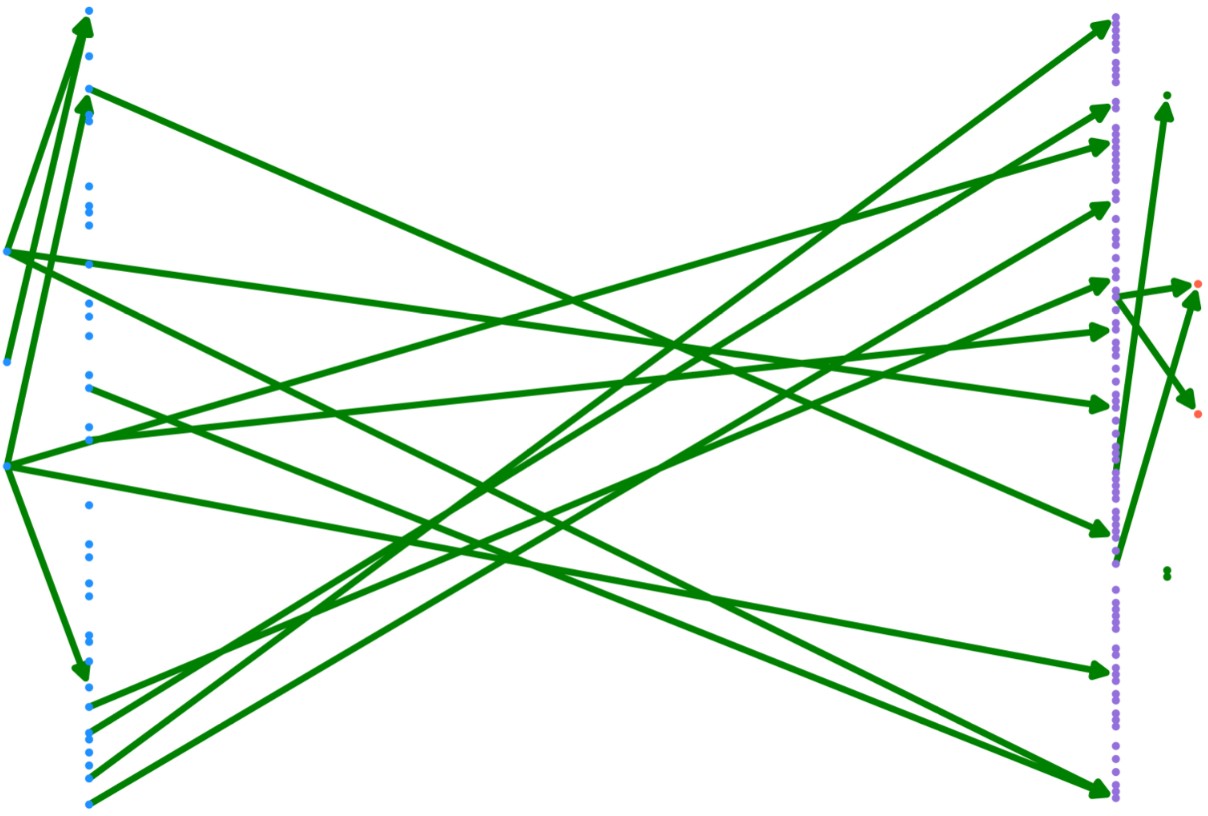}}
\caption{CWE View 1003 Nodes with Direct Edges from Non-1003 Views (123 nodes, 19 edges)}
\label{fig:NewDirectEdges1003CWEGraph}
\end{figure}

We next augment our view 1003 DAG with edges extracted from the `ChildOf' relationships specified within other CWE view XML files. For this we use both the CWE research and development concepts views (essentially alternate taxonomies). We can do this because for our metrics we aren't focused on a particular type of child-parent relationship, we just want to know that a child-parent relationship definitively exists between some pair of CWEs in the view 1003 set. This analysis adds 19 edges, shown in green in Figure \ref{fig:NewDirectEdges1003CWEGraph}. Note that we move three of the class nodes slightly left of the main column of class nodes to enhance visibility because they now have edges to other classes.

\subsection{Indirect Edge Augmentation}
\label{indirectAug}

\begin{figure}
\centerline{\includegraphics[scale=.22]{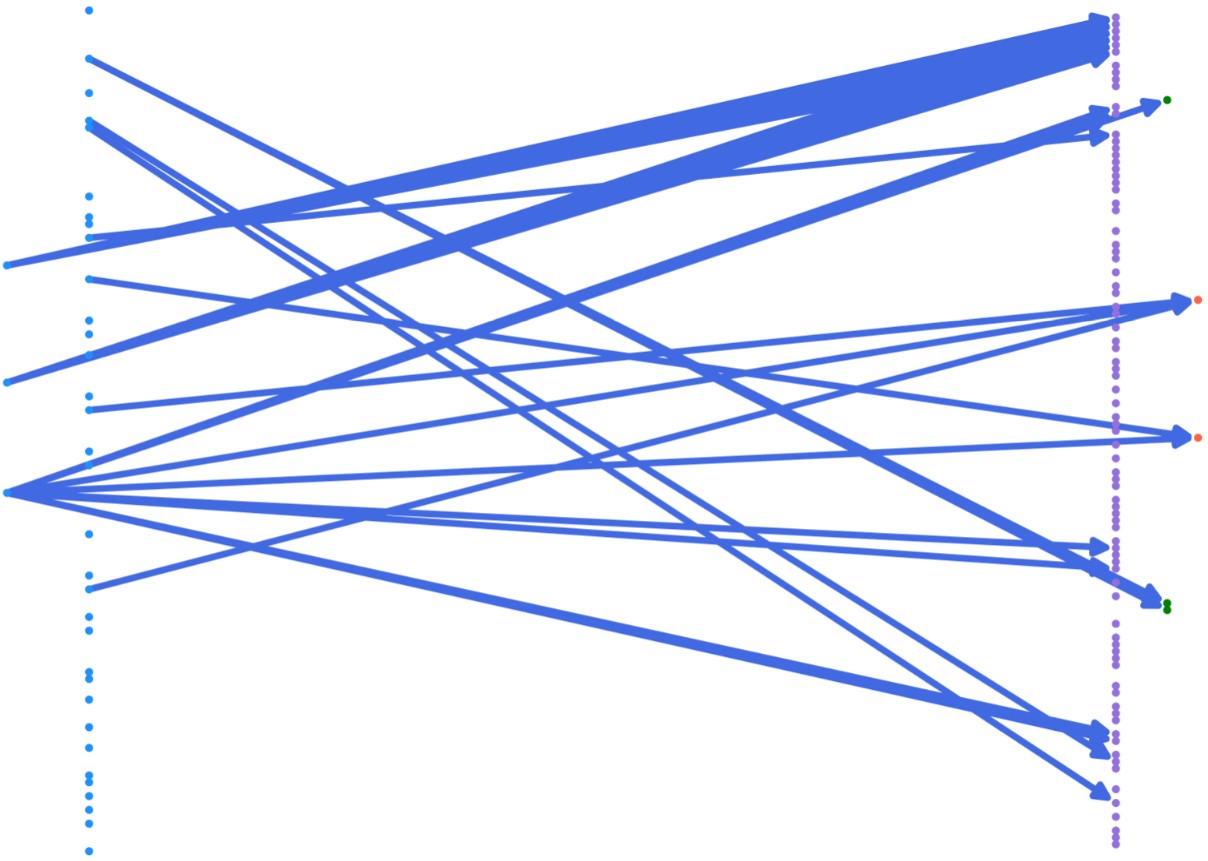}}
\caption{CWE View 1003 Nodes with Edges Representing Paths from Non-1003 Views (123 nodes, 29 edges)}
\label{fig:Path1003CWEGraph}
\end{figure}

Lastly, we create a new DAG (to be used temporarily for this section's analysis) by unifying the set of nodes in views 1003, 1000, and 699 and then adding edges based on the 'ChildOf' relationships specified in the three XML view files. This produces a DAG with 834 CWEs and 1046 edges. Then for each pair of nodes within view 1003, we determine if a path exists connecting them that uses at least one node not in view 1003. Each such discovered path can be used to add an edge to our foundational data structure DAG. These 29 `indirect' edges (that really represent paths using nodes not shown) can be seen in blue in Figure \ref{fig:Path1003CWEGraph}. 

\subsection{Composite Directed Acyclic Graph}

\begin{figure}
\centerline{\includegraphics[scale=.22]{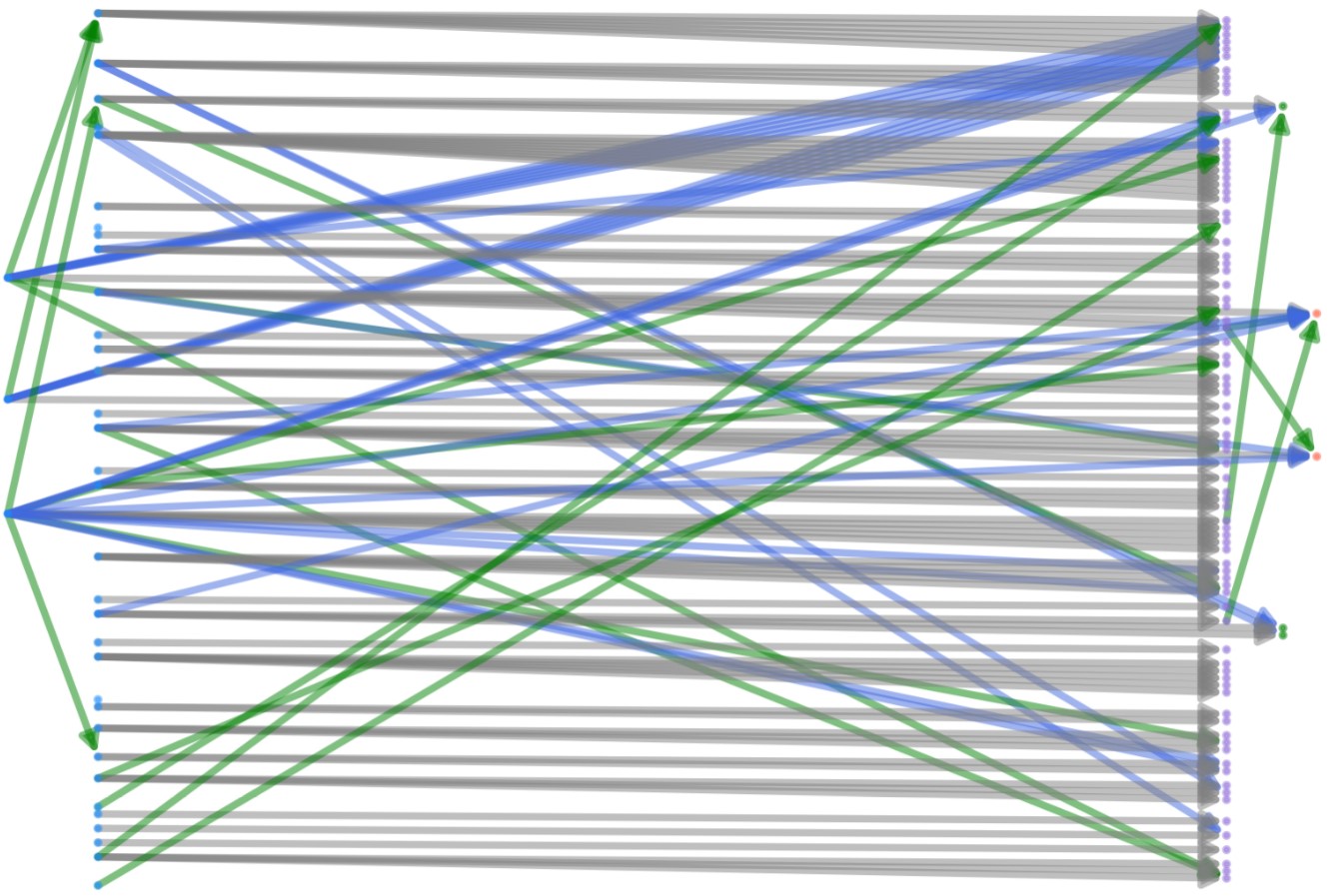}}
\caption{Composite Graph of View 1003 with Direct Edges and Edges Representing Paths from Non-1003 Views (123 nodes, 135 edges)}
\label{fig:Complete1003CWEGraph}
\end{figure}

We now put together our DAG representing the 1003 view with the direct edge augmentation from Section \ref{directAug} and the indirect edge augmentation from Section \ref{indirectAug}. The resulting graph is shown in Figure \ref{fig:Complete1003CWEGraph}. It has 123 nodes and 135 edges.

\subsection{Node Adornment}
\begin{figure}
\centerline{\includegraphics[scale=.22]{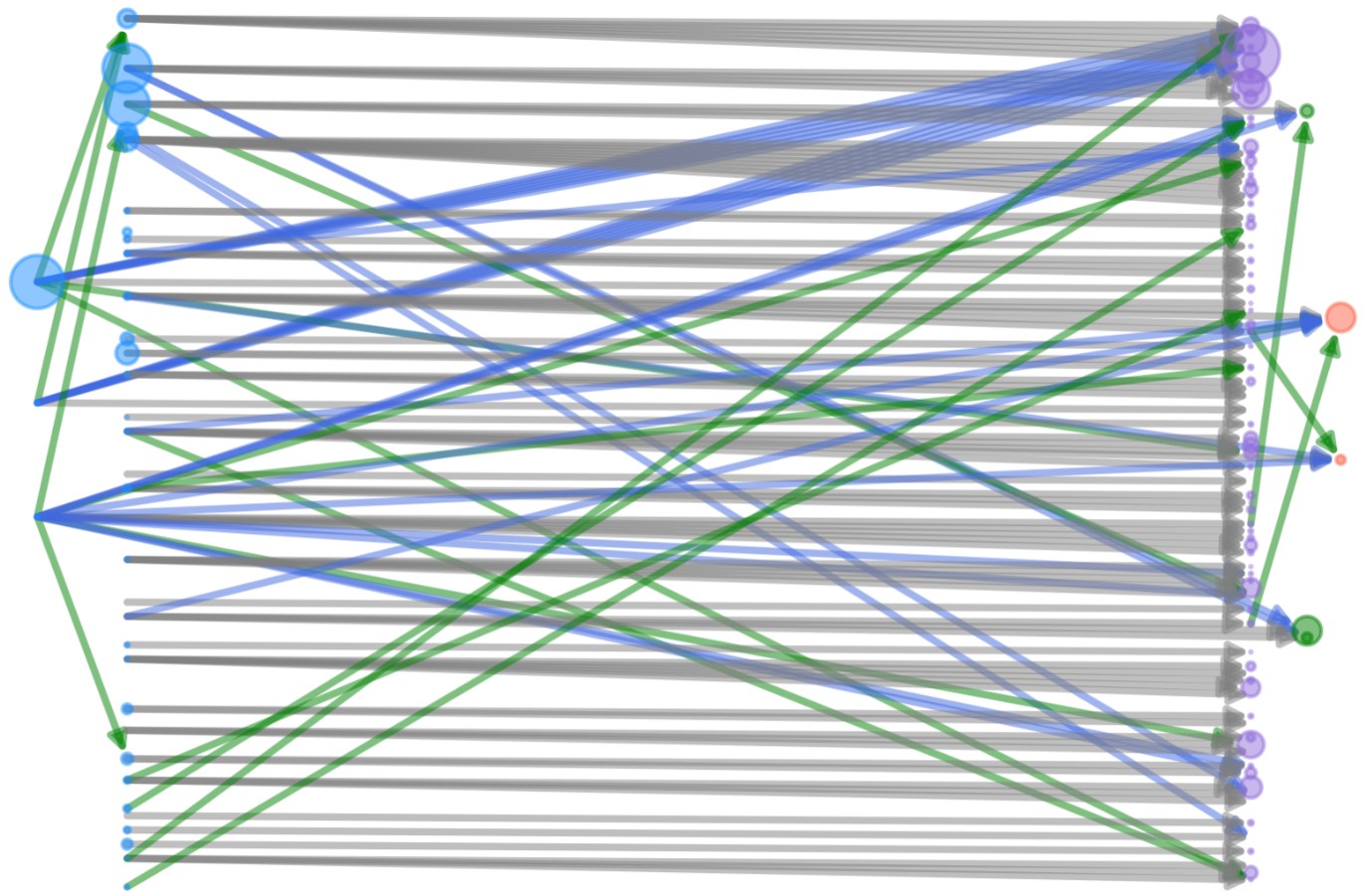}}
\caption{View 1003 Nodes Adorned with NVD Data (no propagation)}
\label{fig:NoProp-CWE-2019}
\end{figure}

The next step is to adorn the DAG with vulnerability analysis data from the NVD. We take each CVE in NVD that has one or more CWE mappings, and we label each relevant CWE node in the DAG with a vector containing the CVE name, the publish date, and the CVSS attribute information. Figure \ref{fig:NoProp-CWE-2019} shows this adornment for the CVEs published in 2019. Note that the size of each node now represents the number of vulnerability vectors mapped to that node.

\subsection{Data Propagation}
\begin{figure}
\centerline{\includegraphics[scale=.22]{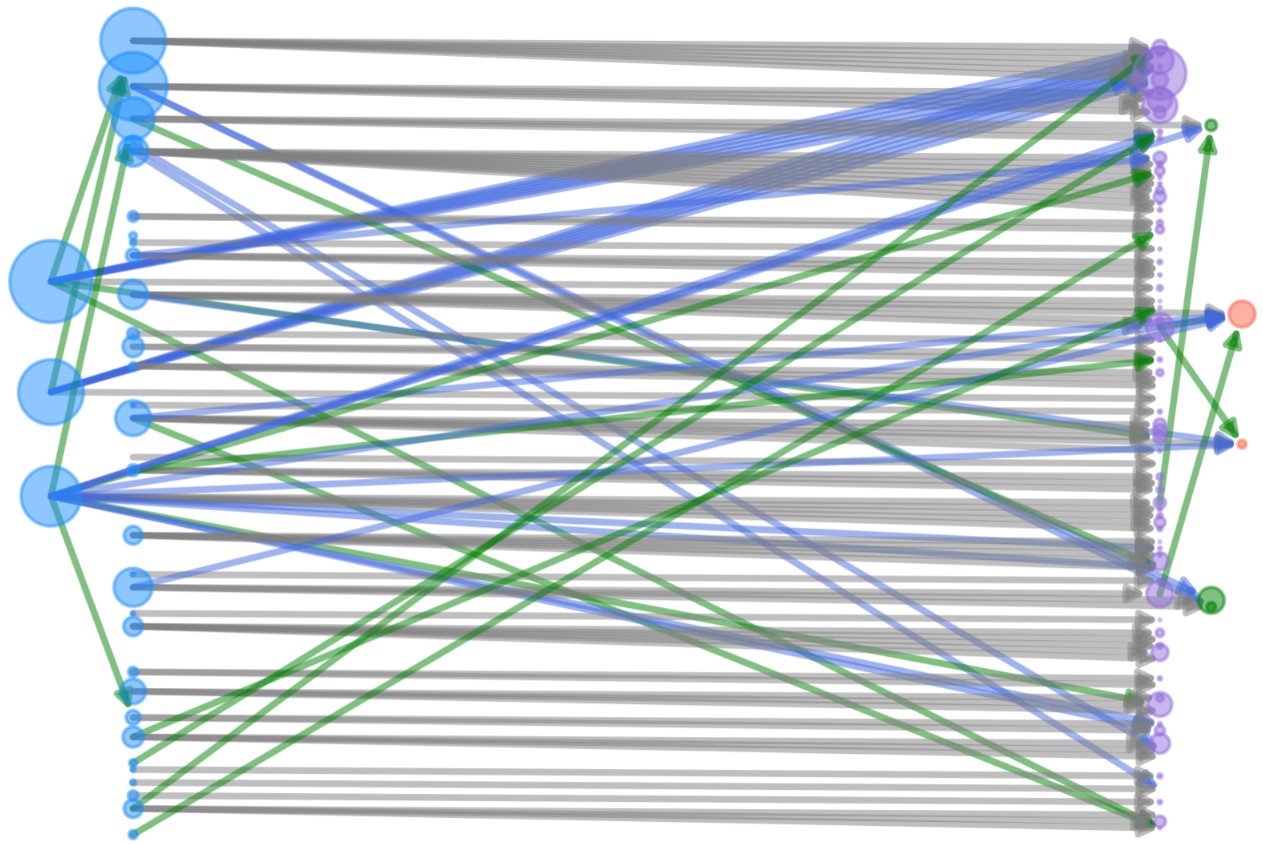}}
\caption{View 1003 Nodes Adorned with NVD Data (with propagation)}
\label{fig:WPropCWE-2019}
\end{figure}

The edges within the DAG represent opportunities for propagating vector data between CWEs. Parent CWEs receive the vectors of their children (with any duplicates being removed). This is because if a vector applies to a CWE then it by definition applies to its more general parent. Also, we discovered that NVD analysts only label a CVE with its most specific CWE. They do not label a CVE with a class if they can determine the applicable base, variant, or compound within that class. Figure \ref{fig:WPropCWE-2019} shows the DAG adorned with the 2019 vulnerability vectors propagated from children to parents. Note, by comparing figures \ref{fig:NoProp-CWE-2019} and \ref{fig:WPropCWE-2019}, how without the propagation some classes get under counted (especially those with many popular bases that are the class' children).

\section{Metrics for Calculating Significance}
\label{Metrics}

The DAG in Figure \ref{fig:WPropCWE-2019} is what we use as the foundational data structure to calculate three simple metrics. The metrics we calculate on this DAG are normalized frequency, mean exploitability, and mean impact. These three metrics are defined below.


We start with a metric to count the number of CVEs mapped to each CWE. Let $I$ designate the set of all CWEs and let $J$ be the set of all CVEs. For CWE $i\in I$, let $N_i$ be the number of CVEs mapped to $i$. We can write it as:

\begin{equation}
N_i = \sum_{j\in J} e_{ij},
\label{eq:N_i}
\end{equation}
where 

\begin{equation}
e_{ij} =
    \begin{cases}
          1, &         \text{if CVE $j$ is mapped to CWE $i$},\\
            0, &         \text{otherwise}.
    \end{cases}
    \label{eq:eij}
\end{equation}

\subsection{Metric for normalized frequency ($F_i$)}

\begin{equation}
F_i = \frac{N_i - \min\limits_{i' \in I} (N_{i'})}{\max \limits_{i' \in I} (N_{i'}) - \min\limits_{i' \in I} (N_{i'})}.
\label{F_i}
\end{equation}

\subsection{Metric for mean exploitability ($Q_i$)}

Let $q_j$ be the CVSS exploitability score for CVE $j$. We can write the average of the $q_j$ in all CVEs mapped to CWE $i$ as:

\begin{equation}
\overline{Q_i} =
\frac{\sum_{j\in J}q_j e_{ij}}{N_i}.
\label{Q_i}
\end{equation}

\subsection{Metric for mean impact ($R_i$)}

Let $r_j$ be the CVSS impact score  CVE $j$. We can write the average of the $r_j$ in all CVEs mapped to CWE $i$ as:

\begin{equation}
\overline{R_i} =
\frac{\sum_{j \in J}r_j e_{ij}}{N_i}.
\label{R_i}
\end{equation}

\section{Weaknesses in the Official CWE Equation}

In September 2019, the official CWE website provided a metric for measuring the `CWE Top 25 Most Dangerous Software Errors' \cite{CWETop25}. It is an aggregate metric combining the normalized frequency of CVEs mapped to CWEs while using the CVSS severity calculated for each mapped CVE.

This metric, like ours, combines together CWE and NVD data, evaluating only the CWEs within the 1003 view. Differing from ours, it uses the raw NVD mappings (it doesn't perform any data propagation) and evaluates all CWE types together (i.e., classes, bases, variants, and compounds). The metric is described in \cite{CWETop25}, we summarize it below (leveraging two of our equations from section \ref{Metrics}).

\subsection{Official CWE Metric}

We first need to define the mean CVSS score for some CWE $i$ as $\overline{S_i}$. 
Let $s_j$ be the CVSS base score for CVE $j$. We can write the average of the $s_j$ in all CVEs mapped to CWE $i$ as:

\begin{equation}
\overline{S_i} =
\frac{\sum_{j\in J}s_j e_{ij}}{N_i}.
\label{S_i}
\end{equation}

Now we define the official `most dangerous' CWE score as $D_i$ for some CWE $i$.
Let $F_i$ refer to equation \ref{F_i} and let $c_j$ be the CVSS score for the $j$-th CVE.

\begin{equation}
D_i=F_i*\frac{\overline{S_i} - \min\limits_{j\in J} (c_j)}{\max \limits_{j\in J} (c_j) - \min\limits_{j\in J} (c_j)}*100.
\end{equation}

\subsection{Weakness 1: Undercounting Parent CWEs}
\label{sec.weak1}
Almost all CWE classes in view 1003 have children, as do a few bases. All CWEs that are parents then get under counted because CVEs that apply to them are often assigned to their children. The official CWE metric does not propagate CVE assignments from children to parents. Also, the NVD analysts assign only the most specific CWE to a CVE; they do not include the parents of marked CWEs. This artificially decreases the importance of CWEs that have children when using the official metric. Using the official CWE metric on the 2019 data, each CWE gets assigned a mean of 87.99 CVEs. Using the propagation proposed in this paper to avoid under counting parents, each CWE gets assigned a mean of 294.71 CVEs.

\subsection{Weakness 2: Class Bias}
\label{sec.weak2}
The inclusion of all CWE types in the official metric (i.e., classes, bases, variants, and compounds) causes some classes to be unfairly promoted to being within the top lists. This is because classes are at a much higher level of abstraction and thus more CVEs will apply to them. Using the official metric, we find that there are 8 classes on the 2019 top 25 list (32\,\%) while there are 36 classes out of 123 CWEs in the 1003 view (29\,\%). While a bias is not particularly apparent here, the bias is muted because the classes are under counted due to weakness 1 (above). To correctly isolate and measure weakness 2, we remove weakness 1 by using the official metric but while performing data propagation from children to parents. Our regenerated 2019 official top 25 list then contains 16 classes (64\,\%); here the classes are vastly over represented since only 29\,\% of the CWEs in view 1003 are classes.

\section{Analysis}

Our approach of propagating CVE data over the CWE taxonomies fills in data missing from the official CWE metric approach. This addresses weakness 1 described in Section \ref{sec.weak1}. Our approach of creating separate top lists for the two levels of CWE abstraction addresses weakness 2, described in Section \ref{sec.weak2}. Thus, we argue that our approach improves over the original. The question though is whether these improvements make any difference in the results.

Using our DAG and the three metrics, we calculated the most significant CWEs for 2019 at the two levels of abstraction.
We now evaluate these results to verify that propagating analysis data over our DAG substantially changes the generated most significant software flaw lists. We also verify that our multiple metrics produce substantially different lists. To compare different rankings, we measure their distance using two related metrics: the Kendall's Tau and the Spearman's Footrule \cite{Rankings}. For two rankings $l_1$ and $l_2$, the Kendall's Tau $K(l_1,l_2)$ measures the number of pairs of elements in $l_1$ that are swapped in their relative positions in $l_2$. The Spearman's Footrule $F(l_1,l_2)$ measures the number of adjacent element swaps that would need to be performed in $l_1$ to convert it into $l_2$. It has been proven that $\forall l_1,l_2, K(l_1,l_2) \leq F(l_1,l_2) \leq 2K(l_1,l_2)$ \cite{Rankings}. 

Both approaches require that the rankings be of the same length and contain the same elements. Thus, when comparing rankings we use the full rankings of all CWEs observed in the data as opposed to comparing top $X$ lists where $X$ is some integer (using some $X$ to limit list length usually results in lists that contain at least one distinct CWE). The number of observed class CWEs in our data was 36 and the number of non-class CWEs was 87. We performed an empirical study to determine K() and F() for random lists of these sizes using 100\,000 trials. The results are shown in Table \ref{tab:distrand}.

\begin{table}[H]
\centering
\caption{Distance Between Random Rankings}
\label{tab:distrand}
\begin{tabular}{lll}
& K() & F() \\
Length 36 & 315 &  432 \\
Length 87 & 1870 & 2522 
\end{tabular}
\end{table}

We first verify that propagating data over our DAG substantially changes the rankings. For each of our metrics, we calculate the full ranking using all available CWEs and then compare that against a ranking created using the same metric but without propagating data over our DAG. The results are shown in Table \ref{tab:dist1}. Overall, they show that the lists do change significantly when using the DAG to propagate data. Note that the distances are less for the non-class CWEs which is remarkable because those lists are more than twice as long as the class CWE lists (longer lists in general produce greater distances due to more elements possibly being out of place). However, this can be explained by noting that there are only 4 edges between the non-class CWEs which diminishes the effect of propagating data from children to parents. There are, on the other hand, 135 edges over which data can be propagated to the class CWEs.

\begin{table*}[]
\centering
\caption{Distance Between Top Lists Created Using Raw CWEs Versus Propagating Data Over the Constructed DAG}
\label{tab:dist1}
\begin{tabular}{lllll}
 & \multicolumn{2}{l}{\begin{tabular}[c]{@{}l@{}}Class CWEs\\ (list length of 36)\end{tabular}} & \multicolumn{2}{l}{\begin{tabular}[c]{@{}l@{}}Non-Class CWEs\\ (list length of 87)\end{tabular}} \\
 & K() & F() & K() & F() \\
Normalized Frequency & 213 & 324 & 115 & 228 \\
Mean Exploitability & 188 & 264 & 81 & 154 \\
Mean Impact & 175 & 256 & 82 & 164 \\
\end{tabular}%
\end{table*}

We next verify that our multiple metrics produce substantially different lists. If this were not the case, then that would argue towards producing just a single list as opposed to multiple lists with different perspectives. Table \ref{tab:dist2} shows the results for the class CWE lists and Table \ref{tab:dist3} shows the results for the non-class CWE lists. Overall, all the lists appear different. The mean exploitability lists have the most distinction from the other two. Comparing the mean exploitability and normalized frequencies lists for the class CWEs we get lists that are even more different than random (see Table \ref{tab:distrand}). Comparing the mean exploitability and the mean impact lists, they are almost as different as the random lists.

\begin{table*}[]
\centering
\caption{Distance Between Top Class CWE Lists Created Using Different Metrics (with Propagating Data on the DAG)}
\label{tab:dist2}
\begin{tabular}{lllllll}
 & \multicolumn{2}{l}{Normalized Frequency} & \multicolumn{2}{l}{Mean Exploitability} & \multicolumn{2}{l}{Mean Impact}\\
 & Kendall Tau & Spearman & Kendall Tau & Spearman & Kendall Tau & Spearman\\
Normalized Frequency & 0 & 0 & 357 & 488 & 213 & 290 \\
Mean Exploitability &  &  & 0 & 0 & 312 & 420 \\
Mean Impact &  &  &  &  & 0 & 0 \\
\end{tabular}%
\end{table*}

\begin{table*}[]
\centering
\caption{Distance Between Top Non-Class Lists Created Using Different Metrics (with Propagating Data on the DAG)}
\label{tab:dist3}
\begin{tabular}{lllllll}
 & \multicolumn{2}{l}{Normalized Frequency} & \multicolumn{2}{l}{Mean Exploitability} & \multicolumn{2}{l}{Mean Impact} \\
 & Kendall Tau & Spearman & Kendall Tau & Spearman & Kendall Tau & Spearman \\
Normalized Frequency & 0 & 0 & 1281 & 1716 & 976 & 1350 \\
Mean Exploitability &  &  & 0 & 0 & 1527 & 1984 \\
Mean Impact &  &  &  &  & 0 & 0 \\
\end{tabular}%
\end{table*}


\section{Related Work}

The NVD also provides CWE rating data. This is in the form of two visualizations that show the relative frequency between the observed CWEs per year and another that shows the actual frequency change for the most frequent CWEs \cite{NVDCWE}. This data is incorrect because it miscounts the frequencies of the CWEs that have children in the CWE taxonomies (because NVD only labels a CVE with its most specific CWE and this information is not propagated to its parents). Also, NVD doesn't distinguish between classes and bases/variants/compounds. The classes are larger categories biasing them to be very frequent, crowding out the bases that are less likely simply due to them being more specific.

While indirectly related, in \cite{zhang2011empirical} a study is done on predicting the next software flaw instances based on NVD data. Also, text clustering has been done on NVD data to automatically identify software flaw types \cite{huang2010text}; this could lead to automatically generated taxonomies to which this work could apply.


\section{Conclusion}
\label{conclusion}

The multiple CWE views can be evaluated as hierarchical taxonomies that reveal parent-child relationships between pairs of CWEs. The different perspectives for each view does not invalidate unifying them because in scoring a CWE as to its significance we want to know all applicable CVEs regardless of the particular method used to organize the CWEs hierarchically. View 1003 provides an obvious base taxonomy from which to start as it was designed to cover the CWEs most used by CVEs. However, its perfect tree structure indicates likely missing relationships. We find those relationships through evaluating the primary three CWE taxonomies (one of which we have to discard because it uses non-CWEs for its higher level classifications). We first find direct missing edges and then find indirect edges (those that represent paths traversing non-1003 view CWEs).

The NVD is an ideal data source to analyze the CWEs because it both maps the CVEs to CWEs and also provides CVSS scores for each CVE. We adorned our unified DAG with the CVE information from NVD and propagated that information from children to parents. We then evaluated the DAG with three simple metrics; focusing separately on classes and bases/variants/compounds due to their very different levels of abstraction. We then generated top lists that provide the most significant CWEs relative to a particular perspective and abstraction. We analyzed those lists and discovered significant differences between them. This argues for the usefulness of and need for multiple top lists with different perspectives. 

It is our hope that software developers and creators of software bug finding tools will use our approach to help prioritize finding and eliminating CWEs in their code. We hope in turn that this will help reduce the number and severity of security related flaws in software.

\section*{Acknowledgement}
This work was partially accomplished under the National Institute of Standards and Technology Cooperative
Agreement No.70NANB19H063 with Prometheus Computing, LLC.

